\documentclass[smallextended]{svjour3}
\smartqed
\usepackage{epsfig}
\usepackage{epstopdf}
\usepackage{graphicx}
\usepackage{subfigure}
\usepackage{caption}
\usepackage{enumerate}
\usepackage{comment}

\graphicspath{{./Figs/}}
\usepackage{amsmath}

\begin{document}

\title{Equivalence of One Loop Diagrams in Covariant and Light Front QED Revisited}

\author{Deepesh Bhamre         \and
        Anuradha Misra		   \and
        Vivek Kumar Singh}        
\institute{Deepesh Bhamre \at
              Department of Physics, University of Mumbai \\
              Santacruz (E), Mumbai-98\\
              Tel.: +91-9969020679\\
              \email{deepesh.bhamre@physics.mu.ac.in}
              \and
                         Anuradha Misra \at
              	          Department of Physics, University of Mumbai \\
              	          Santacruz (E), Mumbai-98\\
              	          \email{misra@physics.mu.ac.in}
              	          \and
              	          	       Vivek Kumar Singh \at
              	          		       \email{mathematicalperson@gmail.com}
              	          }  
              	                      	          
\date{Received: date / Accepted: date}

\maketitle

\begin{abstract}

We revisit the proof of equivalence of one loop expressions for fermion self-energy and vertex correction in light-front QED and Covariant QED at the Feynman diagram level and generalize, to all components, the  proof of equivalence for the one loop vertex correction diagram which was presented earlier by us only for the $+$ component of $\Lambda^\mu$. We demonstrate, in the general case also, that the equivalence cannot be established without the third term in the three-term photon propagator in light cone gauge.

\keywords{Light Front, QED }
\PACS{11.15.Bt,12.20.-m}

\end{abstract}

\section{Introduction}
\label{intro}

The issue of equivalence of covariant and light-front Hamiltonian perturbation theory has attracted a lot of attention in the past years \cite{bmcy05}. Equivalence of LFQED \cite{must} and covariant QED at the Feynman diagram level was addressed, for the first time, by us \cite{swat}, where we demonstrated how one can obtain all the propagating as well as instantaneous diagrams of one loop LFQED by performing the $k^-$-integration over the covariant expressions carefully. It was shown that the equivalence cannot be established by performing $k^-$-integration if one uses the commonly used two-term photon propagator in light cone gauge \cite{must}:

\begin{equation}
d_{\mu\nu}(k) = \frac{1}{k^2+i\epsilon}\bigg[-g_{\mu\nu} + \frac{\delta_{\mu+}
k_{\nu}+\delta_{\nu+}k_{\mu}}
{k^+}\bigg],
\label{eq:prop1}
\end{equation}

\noindent because the instantaneous contribution from the longitudinal polarization of the virtual photon cannot be removed. However, if one uses the three-term photon propagator \cite{brodsky01,suzuki03} given by

\begin{equation}
d_{\mu\nu}^\prime(k) = \frac{1}{k^2+i\epsilon}\bigg[-g_{\mu\nu} + \frac{\delta_{\mu+}k_
{\nu}+\delta_{\nu+}k_{\mu}}{k^+}-\frac{k^2\delta_{\mu+}\delta_{\nu+}}{(k^+)^2}
\bigg],
\label{eq:prop2}
\end{equation}     

\noindent then one can show the exact cancellation between the instantaneous contribution from the longitudinal polarization of the virtual photon and the third term of this three-term photon propagator contributed by the transverse polarization of the virtual photon. It has been shown explicitly in \cite{ji} that the sum of both contribution  from the transverse polarization and the longitudinal polarization of the virtual photon is equivalent to the manifestly covariant photon propagator.

In this work, we first revisit the proof of equivalence for one loop fermion self-energy in Section 2 and clarify the issue raised in \cite{manto}. Then, in Section 3, we extend our proof of equivalence of vertex correction $\Lambda^+$ to a general component $\Lambda^\mu$.

\section{Fermion Self-Energy}
\label{selfenergy}

In covariant perturbation theory, the expression for fermion self-energy in the light-front gauge can be rewritten as a sum of three terms \cite{swat}

\begin{equation} \label{se main}
\Sigma(p) = \Sigma_{1}^{(a)}(p)+  \Sigma_{1}^{(b)}(p) + \Sigma_{2}(p)
\end{equation}
where
\begin{equation}
\Sigma_{1}^{(a)} (p) +  \Sigma_{2}(p)= \frac{ie^2}{2m}\int \frac {d^{4}k}{(2\pi)^4}\frac{{\gamma^\mu}{({p\llap/-k\llap/}+m)}{\gamma^\nu}d_{\mu\nu}{(k)}}{[(p-k)^2-m^2+i\epsilon][k^2-{\mu}^2+i\epsilon]}
\end{equation}
on performing the $k^-$ integration reproduces two of the four self-energy diagrams in LFQED (Figs.1(a) and 1(b) of Ref.\cite{must}), i.e. the standard diagram involving two three-point vertices and the diagram involving the four-point vertex corresponding to instantaneous fermion exchange. (Note that the time-component $(x^+)$ is taken to be upwards in all diagrams in this paper as in \cite{must}.) This is achieved by rewriting the fermion momentum as a sum of an on-shell part and an off-shell part. The other two diagrams involving the four-point instantaneous photon exchange vertex are obtained by performing the $k^-$-integration in $\Sigma_{1}^{(b)}(p)$ given by
\begin{equation}
\Sigma_{1}^{(b)}(p) =-\frac{ie^2}{2m}\int\frac{d^{4}k}{(2\pi)^4}\frac{\gamma^\mu (k\llap/^{\prime}_{on}+m){\gamma^\nu}\delta_{\mu+}\delta_{\nu+}k^2}{[(p-k)^2-m^2+i\epsilon][k^2-{\mu}^2+i\epsilon](k^+)^2},
\end{equation}

\noindent which cancels the corresponding contribution from the longitudinal polarization of the virtual photon. As highlighted in Ref. \cite{swat}, this contribution comes from the third term in the photon propagator and hence the equivalence cannot be proved if we neglect the third term in Eq.\ref{eq:prop2}. 

It was pointed out erroneously in Ref.\cite{manto} that our expression for $\Sigma_{1}^{(a)}(p)$ in Eq.(58) of Ref.\cite{swat} involves off-shell momentum of the virtual photon and hence while converting it to on-shell momentum, one can generate the instantaneous photon diagram from this term itself and hence there is no need to add the third term in the photon propagator. However, the expression in Eq.(58) is actually a light front expression in which $k^-$ integration has already been performed using residue theorem and hence it is understood that the photon is on-shell although it is not mentioned explicitly as in the proof of vacuum polarization equivalence. We shall elaborate on this issue in greater detail in a future publication \cite{am2018}.

\section{Vertex Correction}

The standard covariant expression for vertex correction in the light-front gauge is given by
\begin{equation}\label{vc main}
\Lambda^{\mu}(p,p^{\prime},q) = ie^3\int \frac {d^{4}k}{(2\pi)^4}\frac{\gamma^{\alpha}({p\llap/^{\prime}}-{k\llap/}+m)\gamma^{\mu}({p\llap/}-{k\llap/}+m)\gamma^{\beta}d^{\prime}_{\alpha\beta}(k)}{[(p-k)^2-m^2+i\epsilon][(p^{\prime}-k)^2-m^2+i\epsilon][k^2-{\mu}^2+i\epsilon]}
\end{equation}

\noindent To demonstrate that this expression leads to the light front diagrams in Figs. 2 and 3, we split the fermion momenta in on-shell and off-shell parts \\
${(p\llap/-k\llap/)} = k\llap/^\prime_{on}+\frac{\gamma^{+}[(p-k)^2-m^2]}{2(p^{+}-k^{+})}$,
\\${(p\llap/^{\prime}-k\llap/)} = k\llap/^{\prime\prime}_{on}+\frac{\gamma^{+}[(p^{\prime}-k)^2-m^2]}{2(p^{\prime+}-k^{+})}$.
\\It can then be shown, in a straightforward manner, by performing the $k^-$ integration that the one loop vertex correction can be written as
 
\begin{equation}
\Lambda^{\mu}(p,p^{\prime},q)=\Lambda^{\mu (a)}(p,p^{\prime},q)+\Lambda^{\mu (b)}(p,p^{\prime},q)+\Lambda^{\mu (c)}(p,p^{\prime},q)+\Lambda^{\mu (d)}(p,p^{\prime},q)+\Lambda^{\mu (e)}(p,p^{\prime},q)
\end{equation}
where
\begin{equation}
\Lambda^{\mu (a)}(p,p^{\prime},q)=e^3\int\frac{d^{2}k_{\perp}}{(4\pi)^3}\int_{0}^{p^{\prime+}}\frac{dk^{+}}{k^{+}k^{\prime+}k^{\prime\prime+}}\frac{\gamma^{\alpha}({k\llap/^{\prime\prime}_{on}+m)\gamma^{\mu}(k\llap/^{\prime}_{on}+m)\gamma^{\beta}d_{\alpha\beta}(k)}}{(p^{-}-k^{-}_{on}-k^{\prime-}_{on})(p^{-}-q^{-}-k^{-}_{on}-k^{\prime\prime-}_{on})}
\end{equation}

\begin{equation}
\Lambda^{\mu (b)}(p,p^{\prime},q)=-e^3\int\frac{d^{2}k_{\perp}}{(4\pi)^3}\int_{p^{\prime+}}^{p^+}\frac{dk^{+}}{k^{+}k^{\prime+}k^{\prime\prime+}}\frac{\gamma^{\alpha}({k\llap/^{\prime\prime}_{on}+m)\gamma^{\mu}(k\llap/^{\prime}_{on}+m)\gamma^{\beta}d_{\alpha\beta}(k)}}{(p^{-}-k^{-}_{on}-k^{\prime-}_{on})(p^{-}-p^{\prime-}-k^{\prime-}_{on}+k^{\prime\prime-}_{on})}
\end{equation}

\begin{equation}
\Lambda^{\mu (c)}(p,p^{\prime},q)=2e^3\int\frac{d^{2}k_{\perp}}{(4\pi)^3}\int_{p^{\prime+}}^{p^+}\frac{dk^{+}}{(k^{+})^{2}k^{\prime+}k^{\prime\prime+}}\frac{\gamma^{+}(k\llap/^{\prime\prime}_{on}+m)\gamma^{\mu}(k\llap/^{\prime}_{on}+m)\gamma^{+}}{(p^{-}-p^{\prime-}-k^{\prime-}_{on}+k^{\prime\prime-}_{on})}
\end{equation}

\begin{equation}
\Lambda^{\mu (d)}(p,p^{\prime},q)=e^3\int\frac{d^{2}k_{\perp}}{(4\pi)^3}\int_{0}^{p^{\prime+}}\frac{dk^{+}}{k^{+}k^{\prime+}k^{\prime\prime+}}\frac{\gamma^{\alpha}(k\llap/^{\prime\prime}_{on}+m)\gamma^{\mu}\gamma^{+}\gamma^{\beta}d_{\alpha\beta}(k)}{(p^{\prime-}-k^{-}_{on}-k^{\prime\prime-}_{on})}
\end{equation}

\begin{equation}
\Lambda^{\mu (e)}(p,p^{\prime},q)=e^3\int\frac{d^{2}k_{\perp}}{(4\pi)^3}\int_{0}^{p^{+}}\frac{dk^{+}}{k^{+}k^{\prime+}k^{\prime\prime+}}\frac{\gamma^{\alpha}\gamma^{+}\gamma^{\mu}(k\llap/^{\prime}_{on}+m)\gamma^{\beta}d_{\alpha\beta}(k)}{(p^{-}-k^{-}_{on}-k^{\prime-}_{on})}
\end{equation}
$\Lambda^{\mu (a)}(p,p^{\prime},q)$, $\Lambda^{\mu (b)}(p,p^{\prime},q)$ and $\Lambda^{\mu (c)}(p,p^{\prime},q)$ have been obtained in Ref.\cite{swat} for $\Lambda^+$ and the same proof holds for $\Lambda^\mu$ also. $\Lambda^{\mu (d)}(p,p^{\prime},q)$ and $\Lambda^{\mu (e)}(p,p^{\prime},q)$ which were zero for the $+$ component come from the off-shell parts of the two fermion propagators and have been verified by us to be the same as the light front expressions using LFTOPT. The details of the calculation will be given in a future work\cite{am2018}. It should be pointed out that the sixth term with contribution from $p\llap/-p\llap/_{on}$ for both the fermion lines is zero due to Dirac structure of the numerator. As in the case of fermion self-energy, we again find that the diagram involving instantaneous photon exchange arises due to the third term in the photon propagator.
\begin{figure}[h!]
\centering
\includegraphics[scale=0.27]{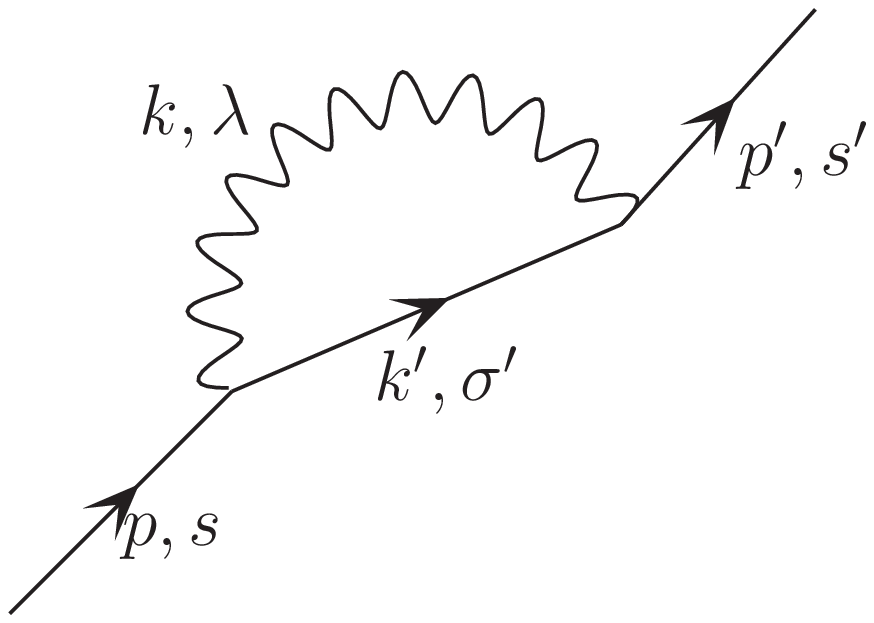}
\includegraphics[scale=0.27]{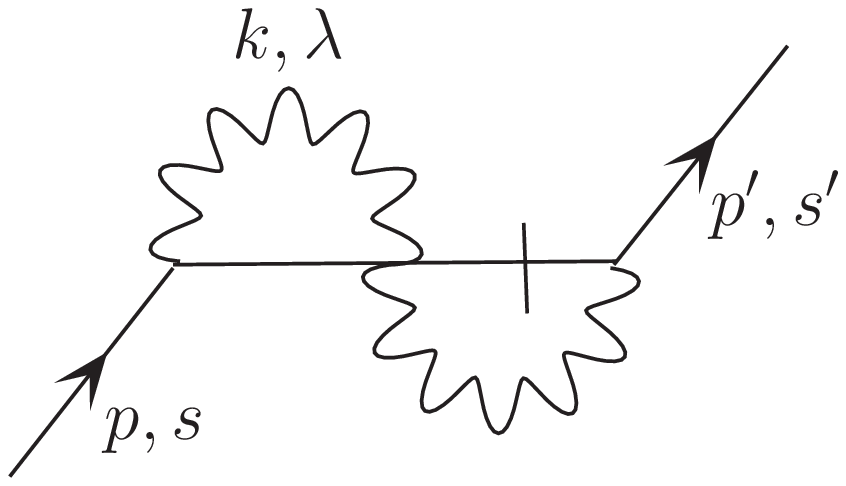}
\includegraphics[scale=0.3]{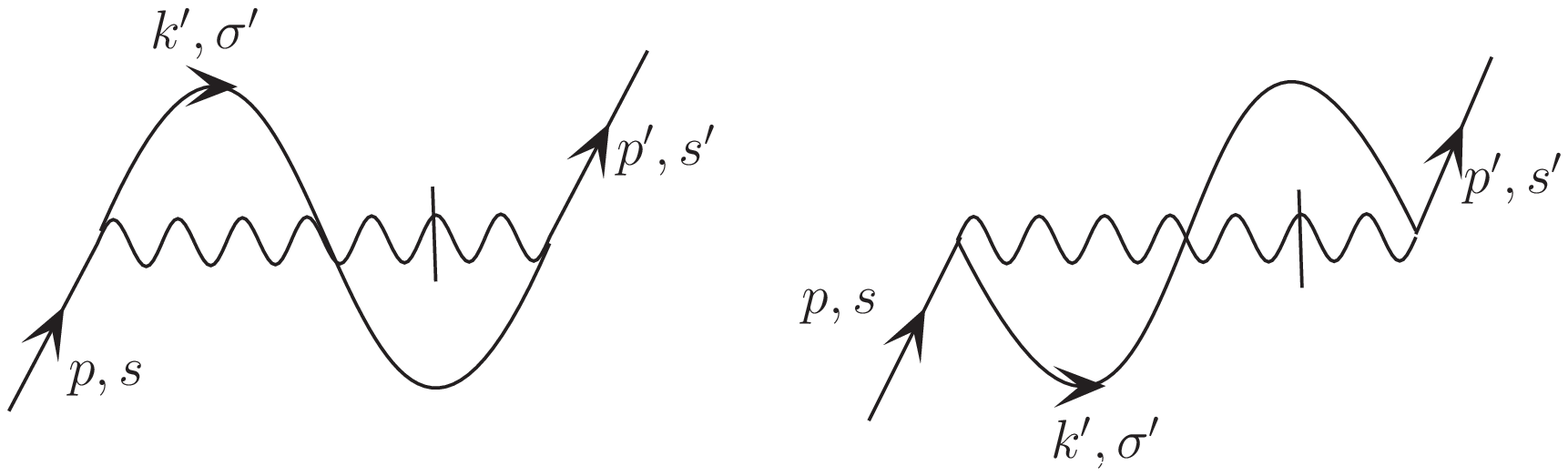}
\caption{Self-Energy diagrams}
\label{sediag}
\end{figure}

\begin{figure}[h!]
\centering
\includegraphics[scale=0.3]{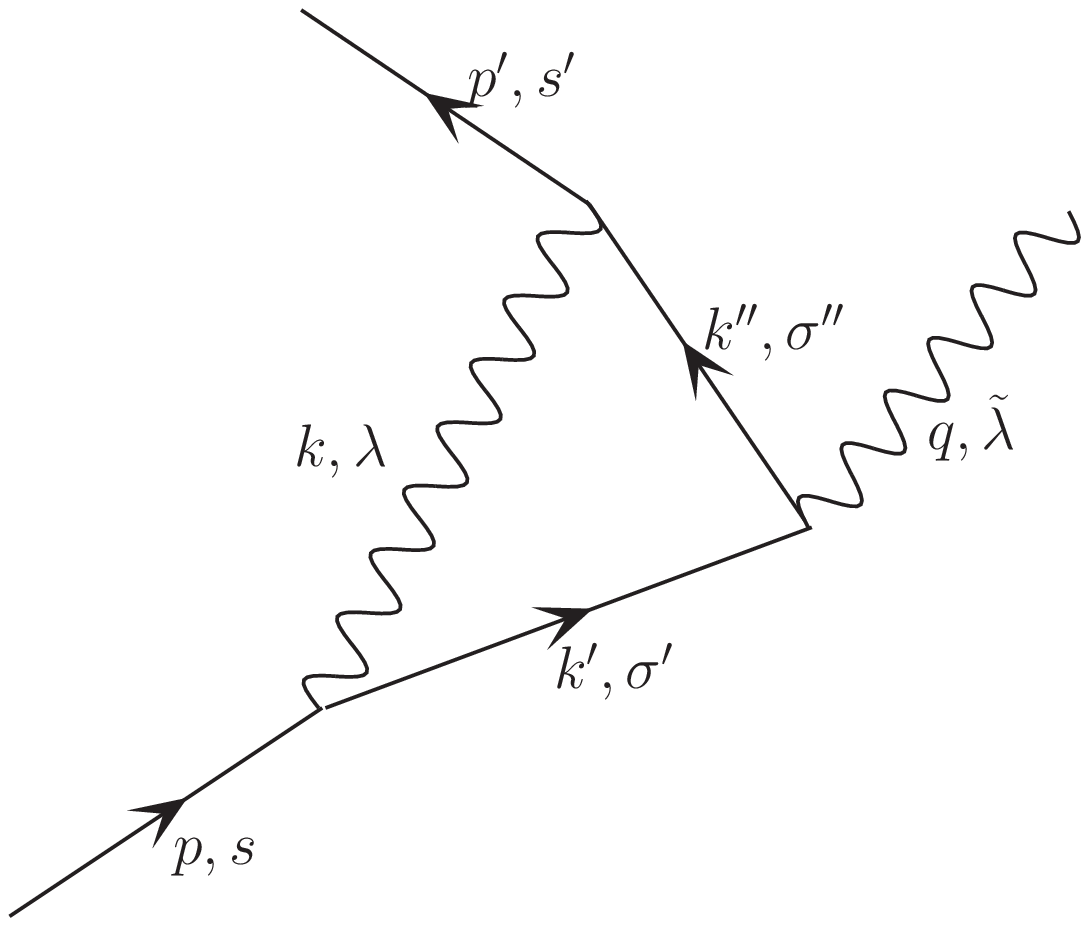}
\includegraphics[scale=0.27]{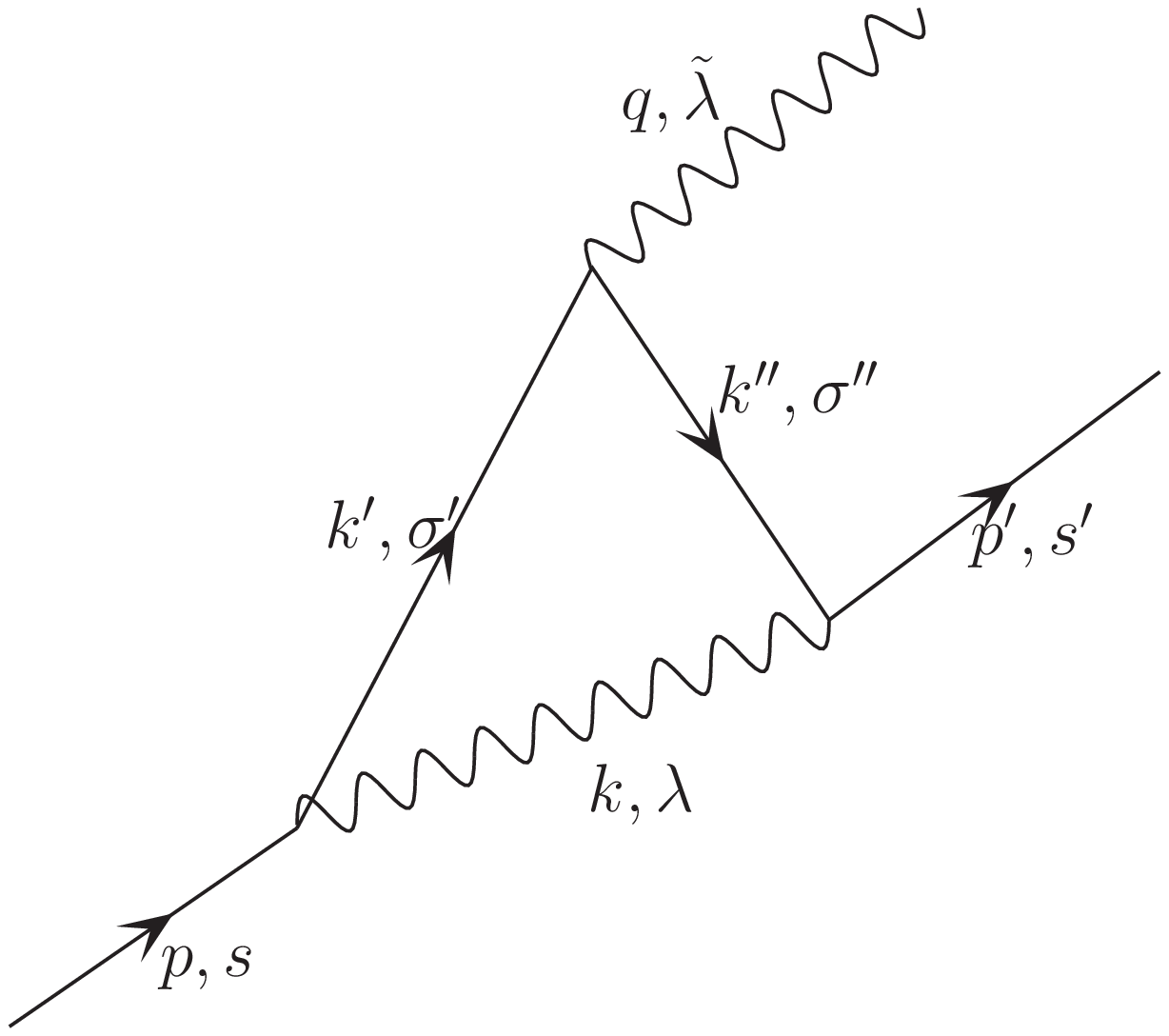}
\includegraphics[scale=0.3]{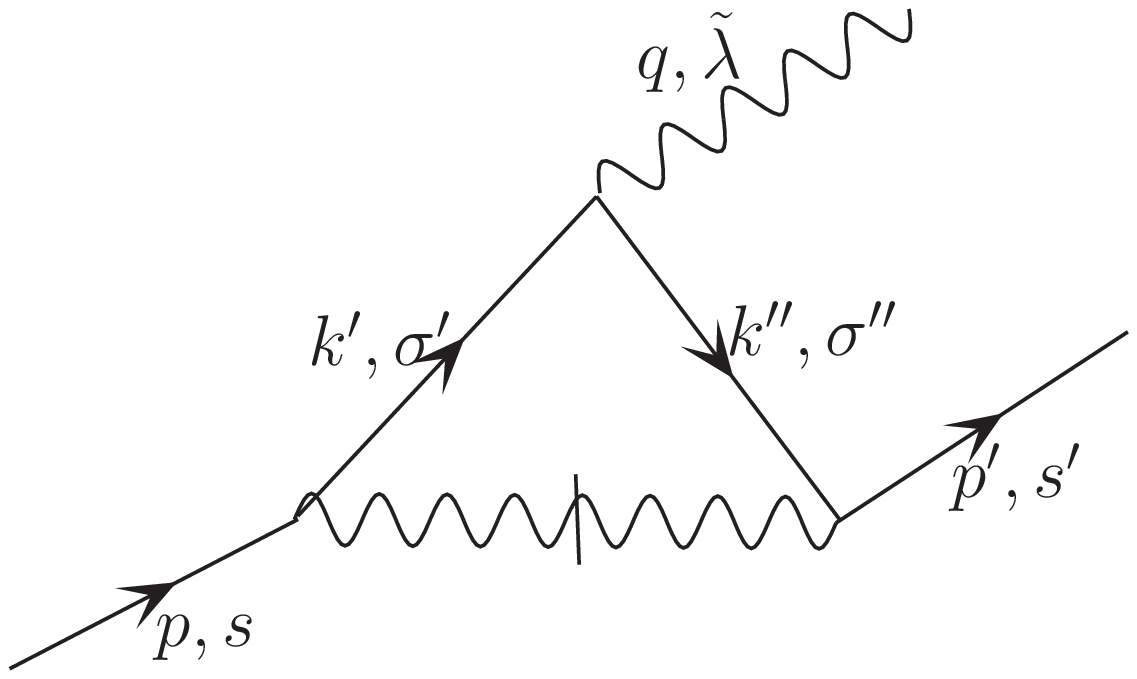}
\caption{"Regular" and instantaneous photon diagrams}
\label{reg inst photon}

\end{figure}
\begin{figure}[h!]
\centering
\includegraphics[scale=0.4]{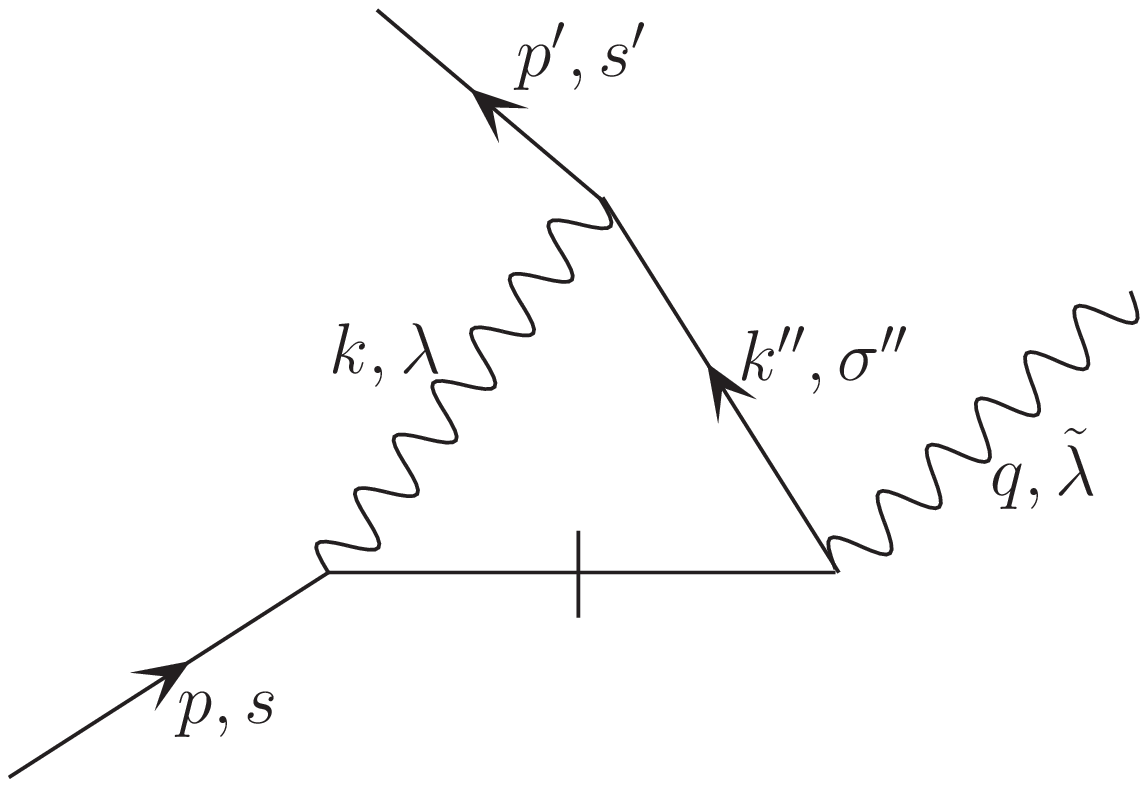}
\includegraphics[scale=0.375]{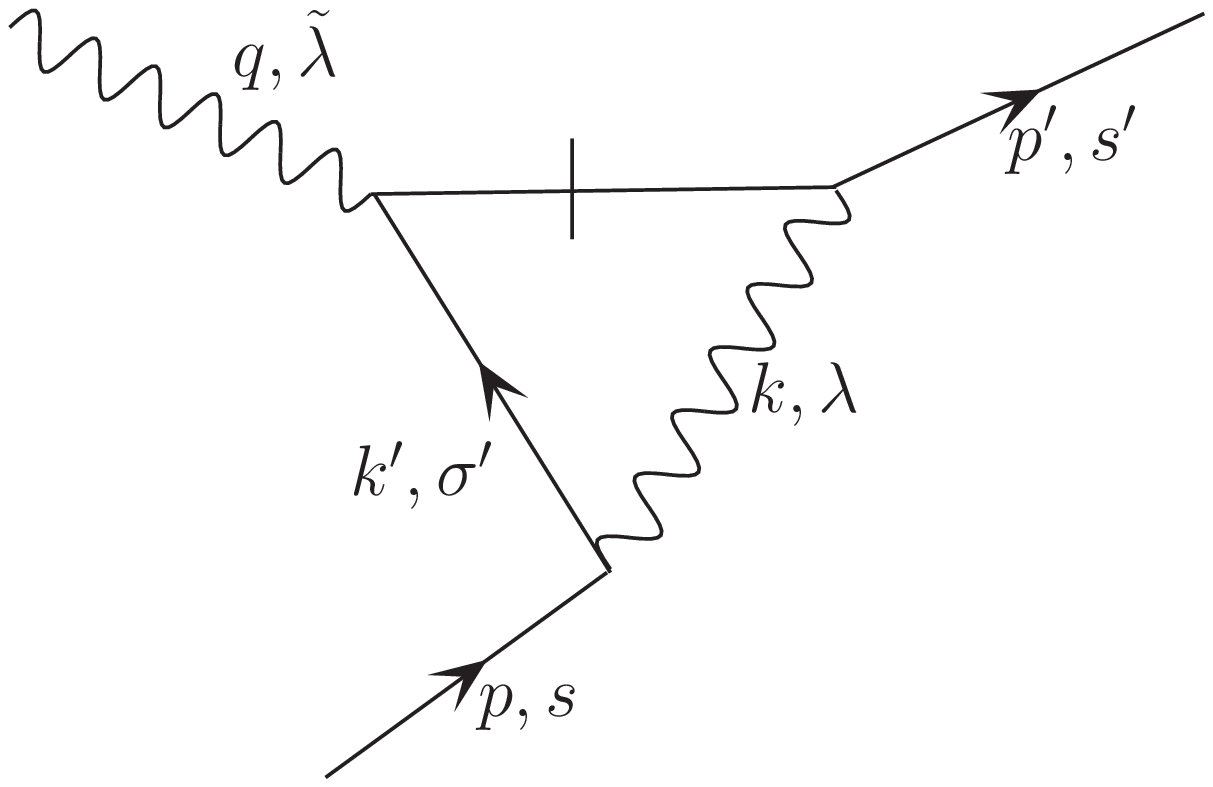}
\caption{Instantaneous fermion diagrams}
\label{inst fermions}
\end{figure}

\vfill

\section{Summary}
We have established equivalence for both the fermion self-energy and the vertex correction contributions, by demonstrating that \\1. The on-shell part of the fermion propagator in the covariant theory, when considered with the two-term photon propagator, is equivalent to the "regular" diagrams in LFTOPT whereas the instantaneous fermion diagrams arise from the off-shell part.
\\2. The third term in the gauge propagator leads to the diagrams containing instantaneous photons and hence should be necessarily considered in order to establish equivalence between covariant and LFQED.

\end{document}